\documentclass[journal=jpclcd,manuscript=article, layout=twocolumn]{achemso}

\usepackage[version=3]{mhchem} 
\usepackage{textcomp}

\author{D\'aniel Datz}
\email{datz.daniel@wigner.hu}
\affiliation[Wigner]
{Wigner Research Centre for Physics, Konkoly Thege Mikl\'os \'ut 29-33., Budapest, Hungary}
\alsoaffiliation[ELTE]
{E\"otv\"os Loránd University, P\'azmány P\'eter s\'et\'any 1/A, Budapest, Hungary.}

\author{Gergely N\'emeth}
\affiliation[Wigner]
{Wigner Research Centre for Physics, Konkoly Thege Mikl\'os \'ut 29-33., Budapest, Hungary}
\alsoaffiliation[BME]
{Budapest University of Technology and Economics, M\H uegyetem rkp. 3., Budapest, Hungary.}

\author{\'Aron Pekker}
\affiliation[Wigner]
{Wigner Research Centre for Physics, Konkoly Thege Mikl\'os \'ut 29-33., Budapest, Hungary}

\author{Katalin Kamar\'as}
\affiliation[Wigner]
{Wigner Research Centre for Physics, Konkoly Thege Mikl\'os \'ut 29-33., Budapest, Hungary}

\title[Title]
  {Polaritonic Enhancement of Near-field Scattering of Small Molecules Encapsulated in Boron Nitride Nanotubes}

\begin{document}
\begin{tocentry}

\includegraphics[width=5cm]{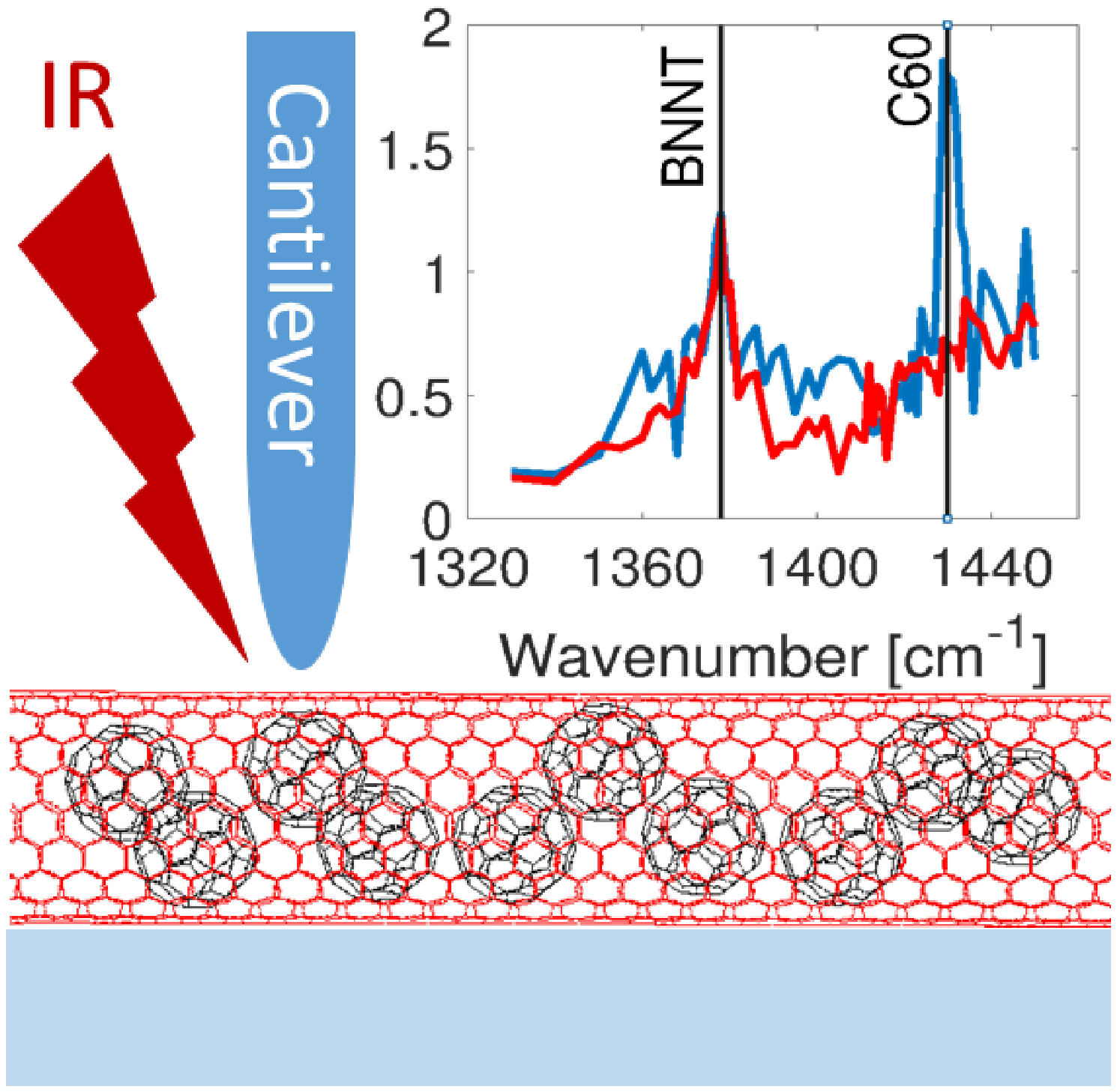}

\end{tocentry}

\begin{abstract}

  Near-field spectroscopy has been extensively applied recently to analyze collective optical properties of materials at the nanoscale. However, vibrations of small molecules were only recognizable in  close proximity to a metallic resonator. We show that encapsulation in boron nitride nanotubes (BNNT) enhances the near-field vibrational spectra of C$_{60}$ fullerene, reaching a sensitivity limit of a few hundred molecules. Furthermore, products of chemical reactions inside the tubes can be identified, so long as their vibrational signatures lie in the polariton gap of the BNNT.

\end{abstract}

Chemical reactions in confined space have been explored with regard to various applications ranging from targeted catalytic reactions \cite{mouarrawis18} to the basic understanding of the evolution of life \cite{grommet20}. The outcome of these reactions is determined by multiple factors including the restricted number of orientations due to steric hindrance and also by the influence of the host system on the enclosed molecules. In this paper we demonstrate the possibility to investigate the products of chemical reactions between encapsulated molecules in nanosized cavities in-situ with 20 nm spatial resolution. This small-scale investigation allows us to study the correlation of chemical reactivity with topology or other variables in the local environment. The method of choice for the measurement is scattering type near-field optical microscopy (s-SNOM) which is a unique tool to analyze nanoscale optical phenomena in a variety of materials.

The setup is based on an atomic force microscope (AFM) that probes the sample with a metallic tip in intermittent contact mode. An infrared laser is focused on the tip with a parabolic mirror that also collects the scattered light from the interacting tip-sample-substrate system. Amplitude and phase information of the scattered light is acquired by the interferometric pseudo-heterodyne method \cite{ocelic06}.

While any excitation that influences the scattered light could be detected by this method, most studies concentrate on collective excitations and their interferences. Advances were made on plasmon-polariton modes in graphene\cite{fei12,gerber14}, phonon-polariton modes in hexagonal boron nitride (hBN) \cite{dai14,li15,li17,zhou17,dai19} and alpha-phase molybdenum trioxide ($\alpha$-MoO$_{3}$) \cite{zheng18} and even the hybrid phonon-plasmon modes of stacked hBN-graphene systems \cite{woessner15}. In a previous publication \cite{nemeth19} we identified clusters of a few hundred metal atoms in single-walled carbon nanotubes (SWNT) based on their free-electron (Drude) absorption. Molecular vibrations remained more elusive, most reports being about polymer films, macromolecules or molecules attached to an antenna surface \cite{aizpurua08, neubrech08, neubrech10, neubrech17,autore18}. Hyperspectral Raman measurements by Gaufr\'es et al. \cite{gaufres16} set a lower detection limit of ten $\alpha$-sexithiophene molecules encapsulated in SWNT, but direct infrared detection at this scale has not yet been performed.

Here we present measurements on few-molecule scattering spectroscopy, specifically on C$_{60}$ fullerene molecules encapsulated in boron nitride nanotubes. Considerable enhancement of the near-field interaction is detected when the molecules are spatially confined in BNNT. The enhancement makes it possible to follow chemical reactions in the confined space, by identifying the products of photopolymerization (dimers and trimers) based on their spectral bands in the polariton gap of the nanotubes.

Layered materials such as hexagonal boron nitride show anisotropy in their dielectric function, thus the dielectric function is different for the in-plane and out-of-plane direction. Hexagonal boron nitride has two vibrational phonon modes in the mid-infrared region. Since BN is a polar material, the transverse and longitudinal vibrations are far apart in frequency, resulting in two polariton gaps (similar to the Reststrahlen band in bulk materials) characterized by  Re$(\epsilon_a)<0$, where $a$ refers to either the in-plane ($\parallel$) or the out-of-plane ($\perp$) component of the dielectric tensor. The anisotropy in hBN results in a scenario where in a given polariton gap only one of these two values is negative, while the other one is positive. The isofrequency curves in such a material are thus hyperbolic as opposed to ellipsoidic. In the upper polariton gap (1360--1610 cm$^{-1}$), Re$(\epsilon_\parallel)<0$ and Re$(\epsilon_\perp)>0$ which describes an open hyperboloid and is referred to as Type-II hyperbolic response. In this region the dielectric function can be described by
\begin{equation}
  \epsilon_a(\omega) = \epsilon_{a,\infty}\left(1 + \frac{\omega_{a,LO}^2 - \omega_{a,TO}^2}{\omega_{a,TO}^2 - \omega^2 - i\omega\gamma_a}\right)
\end{equation}

where $a = \parallel,\perp$ and the parameters are $\epsilon_{\infty}$ = 4.52, $\omega_{LO}$ = 1610 cm$^{-1}$,  $\omega_{TO}$ = 1372 cm$^{-1}$  and   $\gamma$ = 5 cm$^{-1}$. The condition for the enhancement is Re$({\epsilon_{a}})<0$, which is satisfied in the $\left[ \omega_{TO}, \omega_{LO} \right]$ interval. This interval agrees well with the tunability of our laser, and also various molecules -- such as C$_{60}$ -- show vibrational signatures in this range. We treat nanotubes as rolled-up hBN planes, therefore in cylindrical coordinates $\epsilon_r = \epsilon_{\perp}$ and $\epsilon_{\theta} = \epsilon_{z} = \epsilon_{\parallel}$ \cite{zhou17}. We note that Phillips et al. \cite{phillips19} treated similar systems as hollow cylinders with air using effective medium theory. In our case, since the diameters of our tubes are much smaller, the effective medium approximation gives only a negligible correction.

\begin{figure*}
 \centering
 \includegraphics[width=17.0cm]{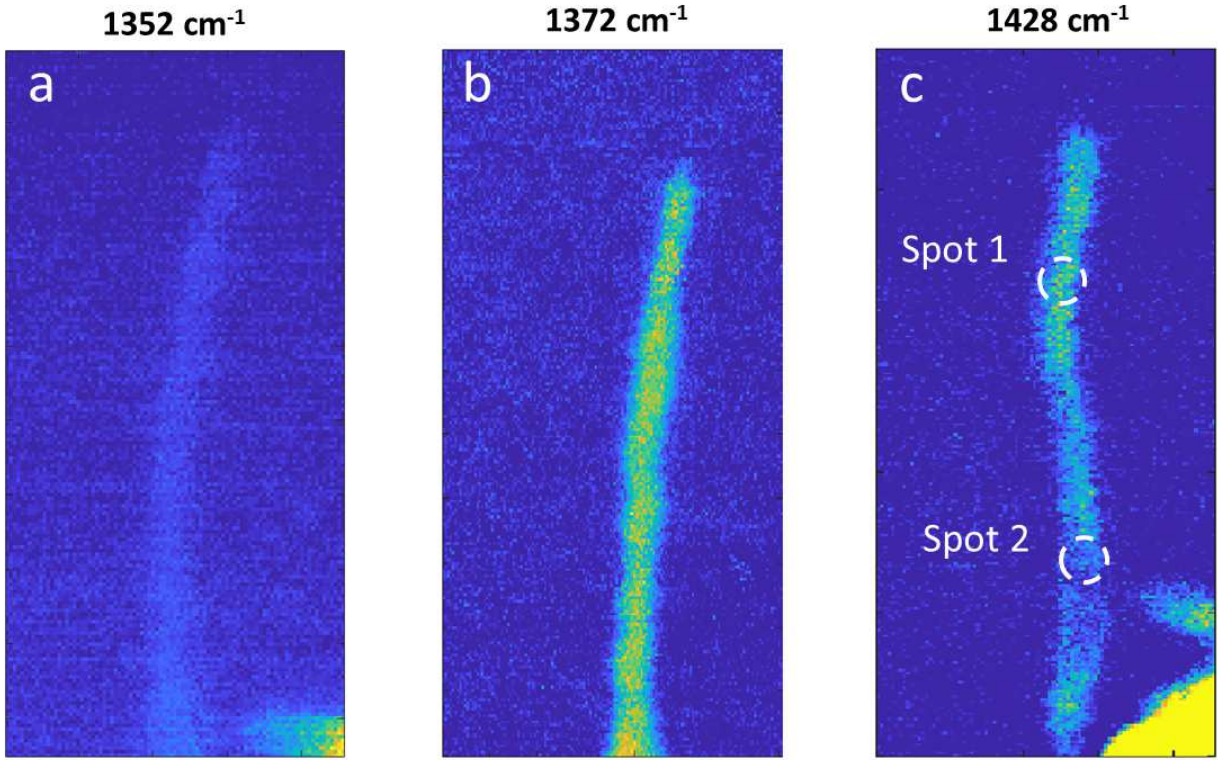}
 \caption{Maps of near-field absorption taken at three different wavenumbers. Nanotube diameter is approximately 3.5 nm. On map c) marked spots indicate the positions where the spectra are calculated (Fig. \ref{fgr:spectra}).}
 \label{fgr:bnntCompFig}
\end{figure*}

Fig. \ref{fgr:bnntCompFig} shows near-field absorption maps of a single, 3.5 nm diameter filled nanotube at three different illumination wavenumbers. At 1352 cm$^{-1}$ no considerable contrast is seen as this wavenumber is below $\omega_{TO}$ and thus below the upper polariton gap. At the TO phonon wavenumber 1372 cm$^{-1}$, the near-field absorption contrast is rather high everywhere along the nanotube, as one would expect. At 1428 cm$^{-1}$, which corresponds to one of the $T_u$ absorption bands of C$_{60}$, high-contrast regions are observed along the nanotube, although less uniform than at the phonon mode frequency (1372 cm$^{-1}$). Since this signal is not present in non-filled nanotubes, we assign it to the encapsulated C$_{60}$ molecular clusters. These results indicate efficient, albeit non-uniform filling.

\begin{figure}[h]
 \centering
 \includegraphics[width=8.5cm]{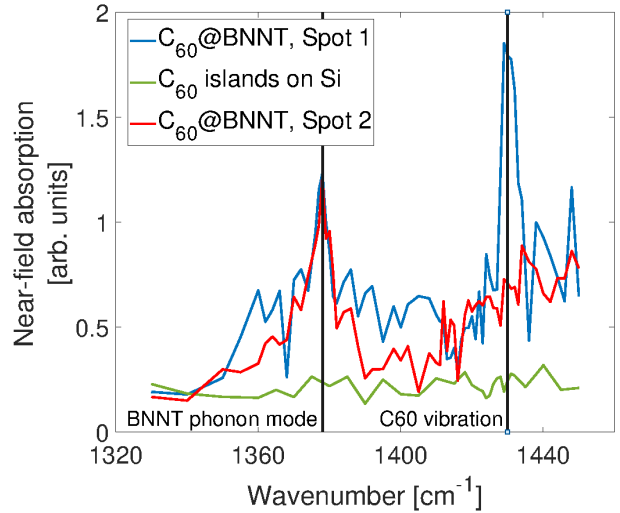}
 \caption{Near-field spectra measured on C$_{60}$@BNNT and C$_{60}$ islands on Si. Spectra were measured at nanotube positions marked by the dashed circles in Fig. 1c. BNNT phonon modes (1372 cm$^{-1}$) and C$_{60}$ modes (1428 cm$^{-1}$) are marked with vertical lines. Comparison with the spectrum measured on C$_{60}$ islands shows the lack of both modes in the absence of BNNT.}
 \label{fgr:spectra}
\end{figure}

As a control experiment, we have also measured C$_{60}$ without the BNNT host. Fig. \ref{fgr:spectra} shows the comparison of the near-field absorption signal of 8--10 nm high C$_{60}$ islands with the filled and unfilled areas of the nanotube of Fig. \ref{fgr:bnntCompFig}. Spectra are acquired by tuning the infrared laser in the range 1330 -- 1450 cm$^{-1}$. The optical signal values from measurement points taken above the silicon substrate are averaged and treated as the reference signal. Measurement points above the nanotube are also averaged and then normalized to the reference signal. Dashed circles in Fig. \ref{fgr:bnntCompFig}c indicate the spots used for extraction of the spectra shown in Fig. \ref{fgr:spectra}. The fullerene absorption at 1428 cm$^{-1}$ is missing from the measurements on the C$_{60}$ islands despite the considerably higher amount of material under the tip. This confirms the amplification of the near-field signal measured in the presence of BNNT in the appropriate wavenumber range (above $\omega_{TO}$).

\begin{figure}[h]
 \centering
 \includegraphics[width=8.5cm]{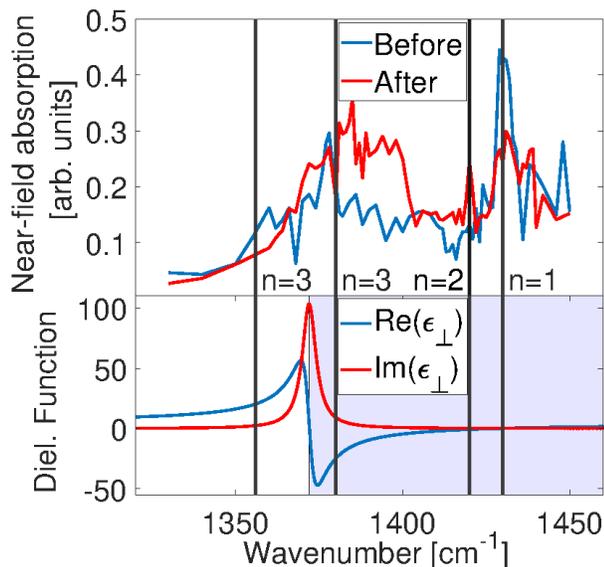}
 \caption{a) Spectra taken on the same section of a C$_{60}$@BNNT before and after intense green laser illumination. New, narrow peaks appear at wavenumbers corresponding to polymerized C$_{60}$ denoted by the vertical lines. b) Boron nitride dielectric function, showing the lower energy side of the upper polariton gap (shaded region). Vertical lines show the vibrational modes of different oligomers of C$_{60}$. The modes in the shaded polariton gap are detectable, while the trimer mode outside is not.}
 \label{fgr:photopolimer}
\end{figure}

In order to demonstrate further applications of the method, we have induced a photopolymerization reaction between the C$_{60}$ molecules using a visible (532 nm) laser focused to the near-field measurement spot. After the initial measurement of the original C$_{60}$@BNNT system we illuminated the same spot with the visible laser for 5 hours, then carried out an identical near-field spectroscopy scan. Upon illumination with the laser, C$_{60}$ molecules undergo photopolymerization resulting mainly in dimers and trimers. The photopolymerization products have distinctive vibrational features that were studied extensively by far field techniques \cite{rao93,rao97,iwasa98,pusztai99,kovats05,klupp07}. Fig. \ref{fgr:photopolimer}a compares results before and after illumination. The C$_{60}$ monomer peak is still visible but an additional mode appears at 1420 cm$^{-1}$ and the peak around 1372 cm$^{-1}$ becomes split and heavily distorted. According  to DFT calculations and measurements from reference \citenum{stepanian06}, these peaks correspond to dimer and trimer absorption modes. The complexity of the higher trimer peak and the nanotube peak may be caused by interaction between the two vibrational modes. It is important to note that the lower trimer absorption band does not appear in our spectrum, as the energy of this vibration is outside of the polariton gap, as seen in Fig. \ref{fgr:photopolimer}b.

The absence of vibrational features below $\omega_{TO}$ confirms the active role of BNNT in the near-field absorption process. To estimate the detection sensitivity obtained by encapsulation, we calculate the volume of the measured region and the number of C$_{60}$ molecules actively participating in the near-field scattering. An upper limit on the number of C$_{60}$ molecules in the measured volume of the nanotube is $\pi r^2_{nanotube}L_{measured}/V_{fullerene} \approx$ 400, where $L_{measured}$ is the length of the nanotube that is in the excited volume of the AFM tip, and is taken as 20 nm. To put this number in perspective, the same amount of C$_{60}$ in a far-field measurement would give an absorption value in the 10$^{-17}$ range \cite{iglesias11}.

In conclusion, we have shown that small-molecule infrared spectroscopy on only hundreds of molecules is possible if the absorption process is supported by the phonon-polariton mode of the encapsulating boron nitride nanotube. The boron nitride cavity also enables observation of few-molecule chemical processes (in this case photopolymerization). The simplicity of the filling process makes filled boron nitride nanotubes appealing candidates for the investigation of the chemical processes in nanotubes.

\section{Experimental}

Nanotubes were purchased commercially (BNNT LLC) and filled with C$_{60}$ using the sublimation method \cite{walker17}. Initially BNNTs were sonicated in 10\% ammonia solution for four hours. The precipitate was filtered through a polytetrafluoroethylene (PTFE) membrane (0.2 $\mu$m) and annealed at 800\textdegree{}C for an hour in air to oxidize the excess boron content. The resulting boron oxide is water soluble and was dissolved in hot water using a vacuum filtration method. The BNNTs were mixed with C$_{60}$ powder in a 1:1 mass ratio, then sealed in a quartz tube at 10$^{-5}$ mbar. The sealed tube was annealed at 600\textdegree{}C for 20 hrs. The resulting powder was thoroughly washed with toluene until C$_{60}$ was undetectable in the UV-Vis spectrum of the filtrate. The filled tubes were dispersed in isopropanol by prolonged, gentle stirring as sonication can force the fullerene molecules out of the nanotubes. Samples were prepared by spincoating the dispersion on a silicon surface (native oxide layer on top).

Scattering near-field measurements were performed with a commercial s-SNOM device (NeaSNOM, Neaspec GmbH) using a quantum cascade laser tunable between 1330--1450 cm$^{-1}$, coupled in and focused on the tip by an off-axis parabolic mirror. The scattered light is collected by a liquid nitrogen cooled mercury cadmium telluride (MCT) detector. Both amplitude and phase response of the scattered light from the sample are measured relative to the substrate. For the photopolymerization, we focused the laser beam to the near-field measurement spot by using the parabolic mirror of the infrared laser.

\begin{acknowledgement}

This research was funded by the Hungarian National Research Fund (OTKA) through grant nos. SNN 118012, PD 121320 and FK 125063. Research infrastructure was provided by the Hungarian Academy of Sciences.

\end{acknowledgement}




\bibliography{Datz_C60BNNT}

\end{document}